\documentstyle[manuscript,aps]{revtex}
\begin{document}
\draft
\title{Fluctuating Bond Aggregation:\\
a Model for Chemical Gel Formation}
\author{R\'emi Jullien and Anwar Hasmy}
\address{Laboratoire de Science des Mat\'eriaux Vitreux,
Universit\'e Montpellier II, Place Eug\`ene Bataillon,
34095 Montpellier, France}
\date{\today}
\maketitle

\begin{abstract}
The Diffusion-Limited Cluster-Cluster Aggregation
(DLCA) model is modified by including cluster
deformations using the {\it bond fluctuation}
algorithm. From 3$d$ computer simulations, it is
shown that, below  a given threshold value $c_g$
of the volumic fraction $c$, the realization of all
intra-aggregate bonding possibilities prevents the
formation of a gelling network. For $c>c_g$, the sol-gel
transition occurs at a time $t_g$ which, in contrast
to DLCA, doesnot diverge with the box size. Several
results are reported including small angle scattering
curves and possible applications are discussed.

\end{abstract}

If the sol-gel process is  widely used to design several materials
of practical interest, the physics of the polymerisation ({\it aggregation})
mechanism that leads to gel formation is far to be completely understood, even
in the
simple case of chemical gels where the  aggregation of $f>2$ functional
monomers is due to polycondensation reactions\cite{1}. Several complementary
theories
have been proposed.
The kinetic approach\cite{2}, which is based on the use of the Smoluchowski
equation,
has the advantage to well describe
the sol-gel transition as due to the appearence of an {\it infinite} aggregate
at a given gel time $t_g$.
But it doesnot give a precise description of the gel structure as all the
geometrical aspects are
hidden in the {\it kernel} of the equation.
On the opposite, the de Gennes and Stauffer approach\cite{3,4}, in which the
sol-gel process is viewed as a
{\it bond percolation} transition\cite{4}, can account for the remarkable
self-similar ({\it fractal})
properties of the gel structure but cannot describe all  the irreversible
kinetical features, as
the percolation theory is restricted to thermodynamical equilibrium.
A great hope was born twelve years ago with the introduction of kinetic
aggregation models\cite{5}
since they were able to describe both  kinetical and geometrical aspects. In
particular the diffusion-limited
cluster-cluster aggregation (DLCA) algorithm\cite{6,7} simulates directly the
aggregation process on a computer
by letting
particles, as well as aggregates of particles, diffusing in a box (with
periodic boundary conditions (PBC))
and sticking irreversibly when they meet.
The DLCA model, as well as some of its variants\cite{8},
were often used  to describe the formation of chemical gels\cite{1,8}.
However, as first pointed out by Kolb and Herrmann\cite{9}, a major problem
arises
as $t_g$ tends to infinity in the thermodynamical limit
of an infinitely large box. Correlatively, the concentration threshold $c_g$,
above which gelation can take place,
vanishes.

In this letter, we show that  allowing cluster deformations during aggregation
removes  these difficulties. Such an extension also provides a more realistic
description
of chemical gel formation, since, in experiments, intra-aggregate motions
always exist up to some degree, due to free-rotations, bond angle deformations
etc...
We report on a $3d$ simulation of an aggregation process of $f$-functional
subunits,
in which, in addition to rigid motions of clusters, some internal movements
are introduced with a tuning {\it flexibility} parameter $F$.
This is made possible by extending the efficient
{\it Bond Fluctuation} method\cite{10,11}.
Such kinetic rules are essentially chosen for the sake of efficiency and it is
clear that
any real dynamics would proceed differently, but
we are convinced that the results obtained are
sufficiently general to have a wide range of applications.
We find that, for a volumic fraction (concentration)
$c$ smaller than a given threshold $c_g$,
the saturation of all intra-cluster bounding possibilities prevents the
formation of a gelling network while for
$c>c_g$, gelation occurs at a given time $t_g$ which does not tend to infinity
with the box size.
Here we give the most significant results
and we show how they can account for some
experiments. A more detailed report will be published later.

Our numerical simulation considers a cubic lattice of unit parameter limited to
a cubic box of edge
length $L$. The moving particles are considered as hard cubes of edge length 2
whose centers can
jump  by one unit from site to site (taking care of PBC). Bonding vectors
between particle centers are restricted to a set
of 108 possibilities:
$[\pm 2,0,0], [\pm 2,\pm 1,0], [\pm 2,\pm 1,\pm 1], [\pm 2,\pm 2,\pm 1], [\pm
3,0,0], [\pm 3,\pm 1,0]$
(and permutations). The directions $[\pm 2,\pm 2, 0]$ (and permutations) are
forbidden to avoid bond crossing\cite{11}.
At $t=0$, $N$ unbonded particles are randomly disposed in the box,
avoiding overlaps (the volumic fraction $c$ is related to $N$ by
$c = 8N/L^3$). Then, we use an iterative procedure in which {\it aggregates}
(= particles connected by bonds), are built.
At a given iteration, an aggregate of $n$ particles (or a single particle if
$n$=1) is
picked up at random with probability $P_n$  and one decides, either to move it
globally and rigidly by a
random unit jump, or to move one of its particles (chosen at random), with
probabilities
$Q_n$ and $1-Q_n$, respectively. If the chosen motion is compatible with bond
restrictions
and hard core conditions, it is performed and one proceeds with the next
iteration.
In the other case, any  overlapping attempt is detected before pursuing with
the next iteration.
If  there is one and if the corresponding
particles are not saturated (i. e. they both have less than $f$ bonds), a new
bond is created
between them (if there are several possibilities, one is chosen at random).
Therefore, in this model, a new bond is formed only when one particle tries to
overlap another one, as in the off-lattice DLCA model \cite{12}.
Another restriction  has been added
which is to forbid the creation of bonds triangles. This artificial (and rather
technical) condition
has been introduced to prevent the formation of tetrahedra
which, due to our bond restrictions, cannot move on large distances by
successive jumps of
single particles.
The following  expressions have been chosen for $P_n$ and $Q_n$:
\begin{eqnarray}
P_n =  {1+Fn\over 1+F}n^\alpha/\sum {1+Fn\over 1+F}n^\alpha; \ \ \ Q_n =
{1\over 1+Fn}
\label{E1}
\end{eqnarray}
where the sum runs over all the clusters and $\alpha$ is a kinetic exponent.
Knowing that, if only single particle jumps are
allowed per unit time, the mobility of an aggregate of $n$ particles  varies as
$1/n$ (as for linear polymers\cite{11}),
such expressions insure, for any $F$, a cluster mobility proportional to
$n^\alpha$, as in DLCA.
It is known\cite{8} that $\alpha$ should be taken equal to $-1/D$,
where $D$ is the cluster fractal dimension, to insure a cluster diffusivity
varying as the inverse of the radius. In all the calculations presented here we
have taken $\alpha = -0.5$, in agreement
with the resulting fractal dimension, of order 2 (we have checked that varying
$\alpha$ around the chosen value
doesnot affect too much the results). As a consequence of (1), the physical
time $t$ is calculated by adding
$\delta t = 1/\sum{1+Fn\over 1+F}n^\alpha$ at each iteration.

We have checked numerically that all the known properties (in particular
$D\simeq 1.8$)
of the 3$d$ DLCA model are well recovered for $F=0$, $f>2$, and in addition,
due to our rule for bond creation, we get no loop,
as in  the off-lattice case\cite{12}.
Here, to illustrate the influence of cluster flexibility, we shall present some
numerical results for a typical large $F$ value
($F=125$).
As the functionality $f$ increases the dynamics is lowered due to  bond
constraints. Therefore, we have taken $f=3$, which is
the lowest value able to produce a gel ($f$=2 corresponds to an interesting
problem of flexible chains aggregation
which has  been previously studied with a different method\cite{13}). The
existence of a  concentration threshold
$c_g\simeq 0.055$ is illustrated qualitatively in fig.1 and quantitatively in
fig.2.
For $c<c_g$, even after waiting a very long time (the figures
correspond to
$t = 10^5\times N$), no gel is obtained and we get a collection of clusters, a
lot of them being
{\it saturated} (i.e. all their particles have $f$-bonds).  In this
concentration range,
we have plotted their mean radius of gyration $R$
as a function of $c$. A more and more pronounced divergence at $c_g$ is
observed as $L$ increases.
For $c>c_g$, we observe that one cluster becomes self-connected via the PBC at
a given time $t_g$, i.e.
it appears a true {\it infinite} cluster when repeating the box by translations
of $L$
in the three directions (note that such condition defines  a gel better than
the one previously used in DLCA\cite{9,12}). The mean values for $t_g$, almost
independent on $L$ (in contrast to DLCA),
also exhibit a quasi-divergence at $c_g$ which is quite well fitted by $t_g\sim
(c-c_g)^{-2}$ (In the case of $R$
we need larger size calculations to be able to estimate any critical exponent).
We have done a few other calculations with different $F$ values and we have
observed that
$c_g$ increases monotically with $F$ and seems to saturate as $F\rightarrow
\infty$ to a limiting
value $c_{g\infty}$ only slightly larger to the one reported here for $F=125$,
while the characteristics
of the transition at $c_g$ remain the same. Moreover preliminary calculations
with $f=4$ indicate that
$c_{g\infty}$ is smaller than with $f=3$.

In figure 3 we report on  numerical results for the Small-Angle Neutron
Scattering (SANS)  curves  of  the gel structure obtained  for  $c>_\sim c_g$,
in the case $f=3$, $F=125$.
 The structure factor $S(q)$, which the Fourier
transform of the particle centers distances, has been calculated by:
\begin{eqnarray}
S(q) = <|\sum_{\overrightarrow r}(\rho(\overrightarrow r)
-\overline\rho)e^{i\overrightarrow q.\overrightarrow r}|^2>
=\sum_{\overrightarrow{r_1}}\sum_{\overrightarrow{r_2}}\bigl(\rho(
\overrightarrow{r_1})\rho(\overrightarrow{r_2})-\overline\rho^2\bigr)
{\sin q|\overrightarrow{r_1}-\overrightarrow{r_2}|\over
q|\overrightarrow{r_1}-\overrightarrow{r_2}|}
\label{E2}
\end{eqnarray}
where $V=L^3$, $\overline\rho = N/V$, $<...>$ designs the average over the
$\overrightarrow q$ directions,
$\rho(\overrightarrow r) =1$ if the site $\overrightarrow r$
is occupied by a particle center and $\rho(\overrightarrow r)=0$ otherwise.
As explained in \cite{12}, it is essential, when one deals with a gel, to
substract  $\overline\rho$ to avoid finite
size artifacts due to the box.  To compare with experiments, it should be
remembered that the SANS intensity curve
$I(q)$ might be obtained from $S(q)$ by multiplying by a form factor
$P(q)$ characterizing the shape of the particles. This has not been done here
due to the crudeness of the model which considers
cubic particles connected by massless bonds.
The log-log curves exhibit the same qualitative shape as for DLCA\cite{12},
i.e. a maximum at low-$q$, which is shifted to larger
$q$ as the concentration increases, an intermediate fractal linear regime,
followed by  damped oscillations
(characteristic of the short range order between particles\cite{14}) at larger
$q$-values. However, for $c>_\sim c_g$,
the absolute value of the fractal slope ($\simeq 2.3$) is
significantly larger than in DLCA  and decreases slowly as $c$ increases. Since
the
fractal dimension of the aggregates forming the gel is very close to 2 (as seen
on the mass-radius curve in inset and also
checked by box-counting calculations), the
slope of $I(q)$ is not equal to $D$ when $c$ is close to $c_g$. This is a
strong indication that the scaling function
characterizing the size-distribution
of clusters should exhibit
a large size power law tail  for $c>_\sim c_g$, $t<_\sim t_g$, in contrast to
the DLCA case
where it is known to decrease exponentially\cite{8}.
Despite numerical uncertainties, a
preliminary calculation of the cluster size distribution  confirms this
hypothesis.
A power law size distribution is not surprising since, in that parameters
range, the system exhibits all the characteristics of
{\it criticality}.
Note also that, for $c>_\sim c_g$, $t<_\sim t_g$, only a few bonds per cluster
are available
to build the gel, and therefore the efficiency
of collisions is very weak as in  Chemically Limited (also called Reaction
Limited) Cluster-Cluster Aggregation for which
it is known that $D\simeq 2$\cite{8}.

In this letter we have shown that including cluster deformations alters
significantly the properties of cluster-cluster
aggregation: a {\it true} sol-gel transition is obtained. More calculations
will be performed to make more precise the roles
of the different parameters, to study more quantitatively the scaling
properties close to $c_g$  and $t_g$, and to compare with percolation theory.
We are convinced that, due to the general character of its results,
the present model can be applied to several
experimental situations. Here
we would like to discuss a possible application to silica aerogels. While base
catalysed aerogels, which result from the
colloidal aggregation of mesoscopic silica particles, are known to form   quite
rigid structures,
well described by DLCA\cite{12}, the neutral  aerogels
are more likely made of a flexible polymeric network, since, from SANS
experiments,
the lower cut-off of the fractal regime is found to be of
atomic scale\cite{15}. The slope of the SANS curve is larger (about 2.3) and a
fractal dimension of $\sim 2.5$ was
reported by fitting  with an analytical formula (valid for a single fractal
cluster)\cite{15}.
Such a larger slope might very well be explained by the present model since it
is also known, from thermoporometry experiments\cite{16},
that the mesoscopic pore size distribution is larger than in base catalysed
aerogels, in agreement with a strong cluster size
polydispersity. We are now suggesting some more experiments to test
such a conjecture.

We benefited from discussions with Marie Foret and Thierry Woignier. One of us
(A.H.) would like
to acknowledge support from CONICIT (Venezuela). Numerical calculations were
done
with the help of the computers of the CNUSC (Centre Universitaire Sud de
Calcul).

\begin{figure}
\caption{Typical configurations obtained with $f=3, L=120, F=125$. Cases (a)
and (b) correspond to $c=0.02$ and $c=0.06$,
respectively. In case (a) the aggregation process was run up to $t=10^5\times
N$ while in case (b) it has been
stopped after obtaining a single cluster. Bonds between particles connected via
periodic boundary conditions are not shown}
\end{figure}
\begin{figure}
\caption{Mean radius of gyration $R$ (circles) and gel time $t_g$ (squares) as
a function
of $c$ for $f=3$, $F=125$ and two box sizes $L=60$ (open symbols) and 120
(filled symbols).
The data result from an average over 50 and 10 independent simulations for
$L=60$ and 120, respectively.}
\end{figure}
\begin{figure}
\caption{SANS  curves $S(q)$ calculated for $f=3$, $L=60$, $F=125$ and three
values of $c$: $c=0.055, 0.065, 0.075$.
For comparison the curve corresponding to $c=0.065$, $F=0$ is shown (dashed
line). In inset the mass-size curves of the
aggregates for $c=0.065$ $F=125$ (solid line) and $F=0$ (dashed line) are
shown.
In the figure and inset the curves have been arbitrarily vertically shifted for
convenience. }

\end{figure}

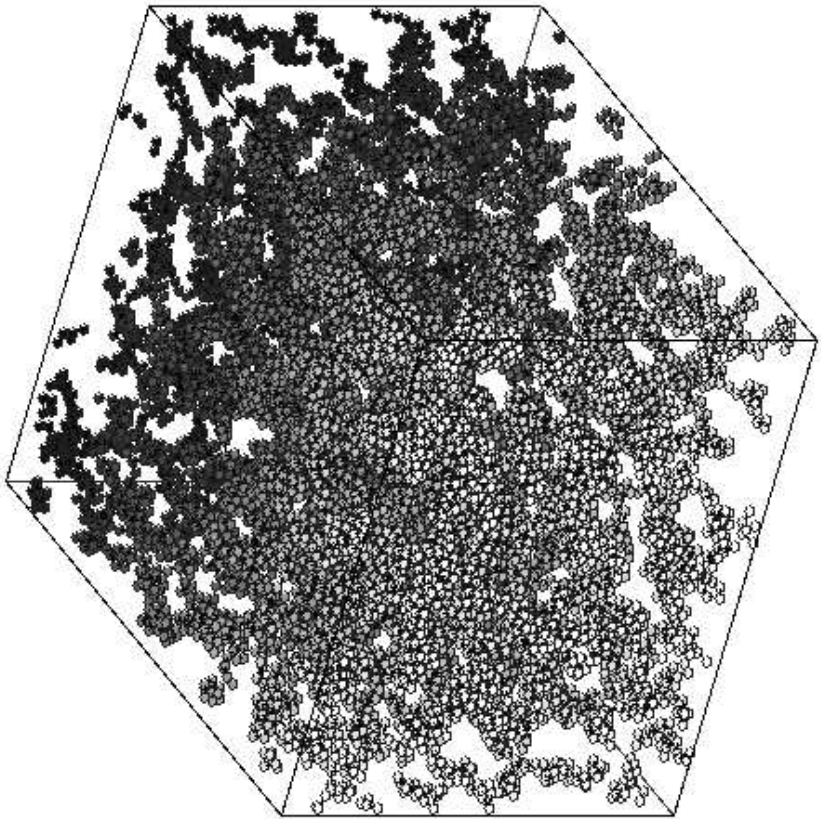
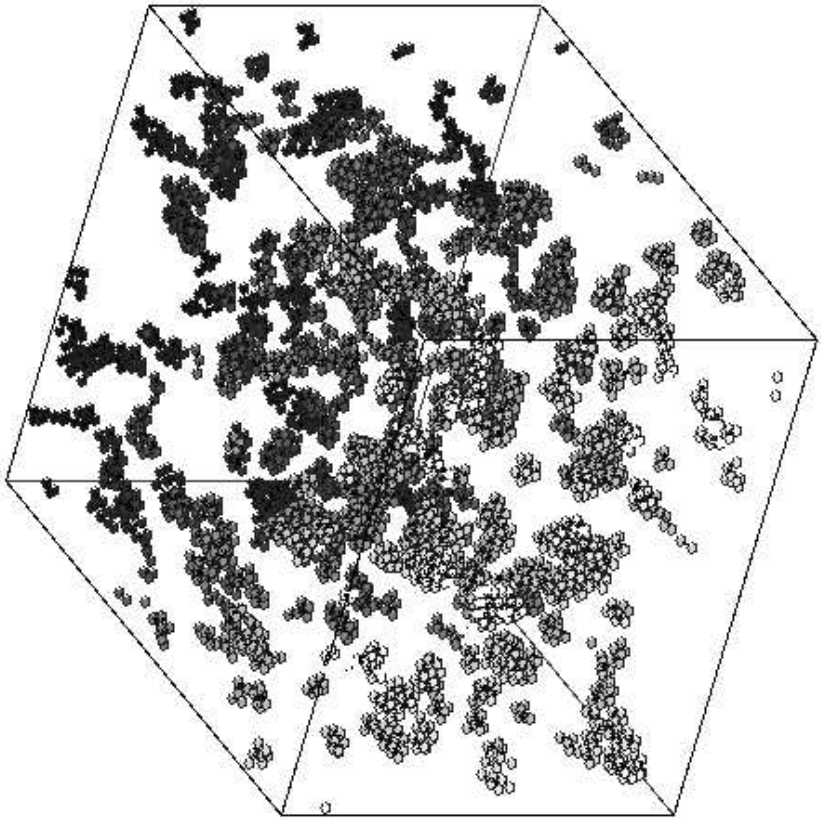
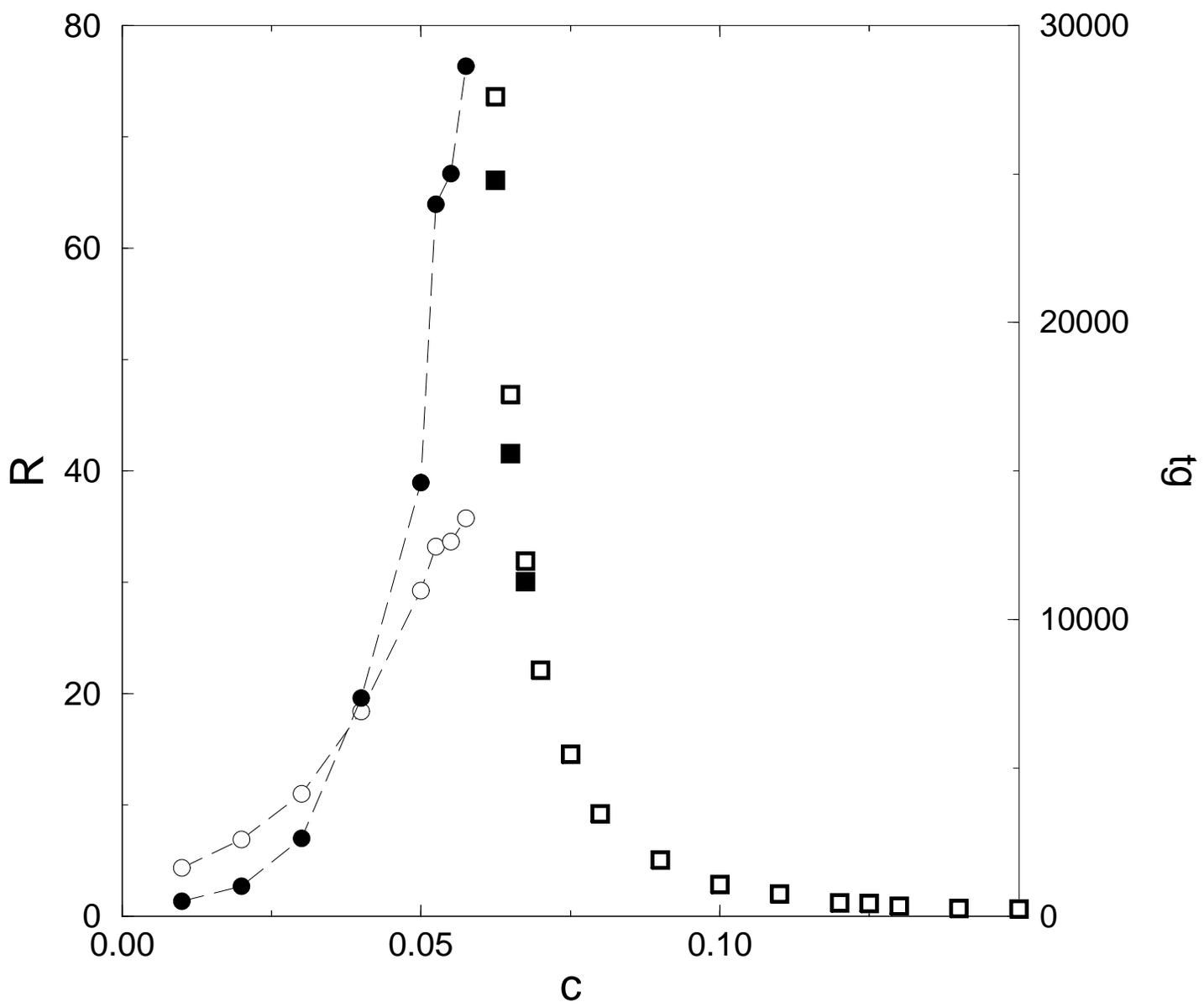
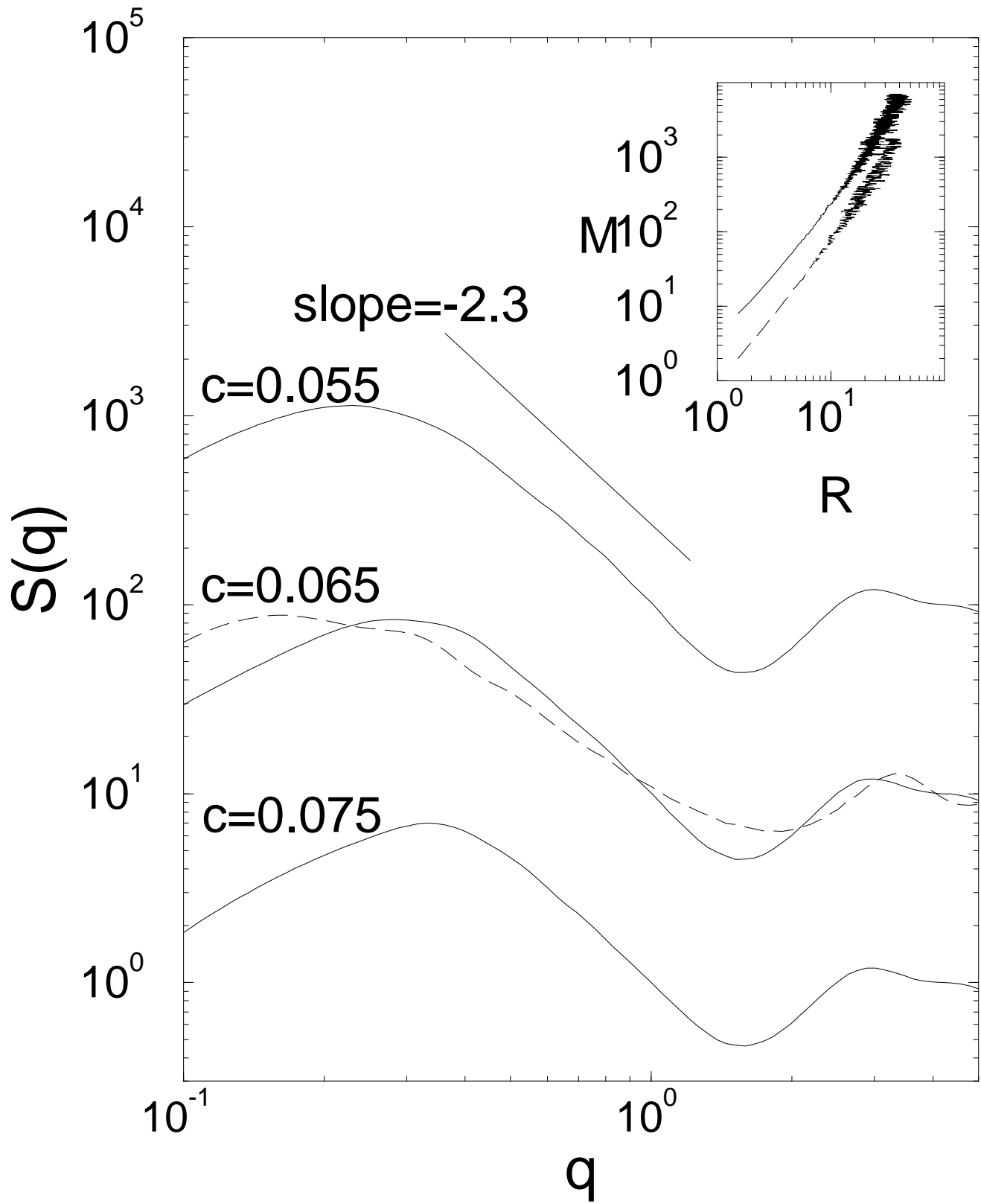
\end{document}